\documentclass[twocolumn,twoside,slac_two]{revtex4}
\usepackage{graphicx}
\usepackage{fancyhdr}
\usepackage{amsmath,amsfonts}
\newcommand\pp{$\text{\textit{p}} \mathop{\text{+}}
  \text{\textit{p}}$}

\newcommand\CuCu{$\text{Cu} \mathop{\text{+}} \text{Cu}$}
\newcommand\AuAu{$\text{Au} \mathop{\text{+}} \text{Au}$}

\newcommand\RAA{$\text{\textit{R}}_{\text{\textit{AA}}}$}

\newcommand\pT{$\text{\textit{p}}_{\text{\textit{T}}}$}
\newcommand\stwohundred{$\sqrt{s} = 200\,\mathrm{GeV}$}
\newcommand\sNNhundredthirty{$\sqrt{s_{NN}} = 130\,\mathrm{GeV}$}
\newcommand\sNNtwohundred{$\sqrt{s_{NN}} = 200\,\mathrm{GeV}$}

\begin{document}

\title{Direct jet reconstruction in \pp{} and \CuCu{} at PHENIX}

%

\author{Y.-s. Lai (for the PHENIX collaboration)}
\affiliation{Columbia University, New York, NY 10027-7061 and Nevis
  Laboratories, Irvington, NY 10533-2508, USA}

\begin{abstract}
  The Relativistic Heavy Ion Collider collides heavy nuclei at
  ultrarelativistic energies, creating a strongly interacting,
  partonic medium that is opaque to the passage of high energy quarks
  and gluons. Direct jet reconstruction applied to these collision
  systems provides a crucial constraint on the mechanism for in-medium
  parton energy loss and jet-medium interactions. However, traditional
  jet reconstruction algorithm operating in the large soft background
  at RHIC give rise to fake jets well above the intrinsic production
  rate of high-\pT{} partons, impeding the detection of the low cross
  section jet signal at RHIC energies. We developed a new jet
  reconstruction algorithm that uses a Gaussian filter to locate and
  reconstruct the jet energy. This algorithm is combined with a fake
  jet rejection scheme that provides efficient jet reconstruction with
  acceptable fake rate in a background environment up to the central
  $\mathrm{Au} + \mathrm{Au}$ collision at \sNNtwohundred{}. We
  present results of its application in \pp{} and \CuCu{} collisions
  using data from the PHENIX detector, namely \pp{} cross section,
  \CuCu{} jet yields, the \CuCu{} nuclear modification factor, and
  \CuCu{} jet-jet azimuthal correlation.
\end{abstract}

\maketitle

\thispagestyle{fancy}


\section{Introduction}

Measurements of single particle production at the Relativistic Heavy
Ion Collider (RHIC) have found significant suppression in \CuCu{} and
\AuAu{} collisions at \sNNhundredthirty{} and \sNNtwohundred{} (e.g.\
\cite{Adcox:2001jp,Adare:2008cx}). However, single or few-particle
observables are fragmentation dependent, and only indirectly probe the
energy loss of the parent parton. While direct jet reconstruction has
become widely used to study perturbative quantum chromodynamics (PQCD)
in $e^+ e^-$ and hadronic colliders, the high multiplicity and
strongly fluctuating background in heavy ion collisions made direct
applications of jet reconstruction difficult.

In PHENIX, the limited detector aperture presents additional
difficulty to directly apply traditional jet reconstruction algorithms
that are sensitive to large angle fragments (or the lack of such),
such as the cone~\cite{Huth:1990mi} and $k_\perp$
algorithms~\cite{Catani:1992zp,Ellis:1993tq}. PHENIX has two ``central
arm'' spectrometers with an aperture consisting of two ``arms'' with
$|\eta| < 0.35$ and $\Delta\phi = \pi/2$ each~\cite{Adcox:2003pd}.
Since PHENIX is designed to measure rare and electromagnetic probes,
and therefore can sustain a high read-out rate, it is well-suited to
measure high-\pT{} jets at a low cross section. However, the narrow
PHENIX central arm aperture especially in $\Delta\eta$ leads to the
loss of large angle fragments from jets and potentially larger
systematic errors due to the importance of edge effects. With
increasing $p_T$, the fragments of jets become increasingly
collimated, with most of the energy concentrated in a cone much
smaller than the PHENIX acceptance. A jet reconstruction algorithm
that emphasizes the core and deemphasizes the large angle tail would
therefore be less sensitive to both the background fluctuation in
heavy ion events and the limited aperture of the PHENIX detector.

In order to provide an effective method to reconstruct jets at the
presence of heavy ion background and/or limited detector aperture, we
started in 2006 to develop a jet reconstruction algorithm that takes
advantage of the collimated emission of hadrons and is insensitive to
large angle fragments. We observed that the flat weighting in
traditional jet reconstruction algorithms makes it prone to large
angle fluctuations. While the background grows with $R^2$, the jet
contribution to the $p_T$ grows slowly above $R \gtrsim 0.3$ (e.g.\
\cite{Acosta:2005ix}). A nonflat weighting that enhance the
core jet signal to periphery would naturally suppress this sensitivity
and stabilize the jet axis. The energy flow
variable~\cite{Berger:2003iw} also provided us with the hint that
angular convolution of the event transverse momentum (\pT{}) density
with a continuous distribution can be an effective description of QCD
processes. We are therefore using a Gaussian filter as a generalized
form of the cone algorithm to reconstruct jets
(\cite{Lai:2008zp,Lai:2009zq}). Early in the development of jet
reconstruction algorithms, the British-French-Scandinavian
collaboration used the equivalent to a Gaussian filter with $\sigma =
0.5$~\cite{Albrow:1979yc}, preceding even the Snowmass accord on the
cone algorithm~\cite{Huth:1990mi}.

The heavy ion background itself is generated by combination of soft,
collective processes, semihard processes and subsequent QCD
hadronization. The presence of heavy flavor decays and minijet
production suggest that one should expect the background to contain
more complex, angularly correlated structures than from a purely
randomized and isotropic process. Therefore, a strategy based on
statistically subtracting the background would require detailed
knowledge about the structure of the heavy ion background and its
interaction with the jet reconstruction algorithm.

\begin{figure}[b]
  \centerline{\includegraphics[width=2in]{%
      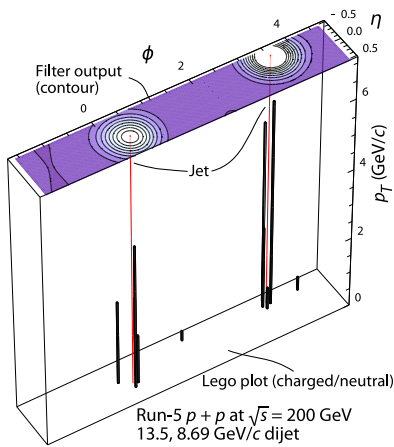}}
  \caption{A PHENIX Run-5 \pp{} at \stwohundred{} dijet event. Charged
    tracks and photons are shown at the bottom by a Lego plot. The
    distribution of filter output values of the event is shown at the
    top as a contour plot. The maxima in the filter density are
    reconstructed as jet axes, shown as red lines at the positions on
    the contour and Lego plots.}
  \label{fig:run_5_p_p_event_display_back_0}
\end{figure}

\begin{figure}[b]
  \centerline{\includegraphics[width=2in]{%
      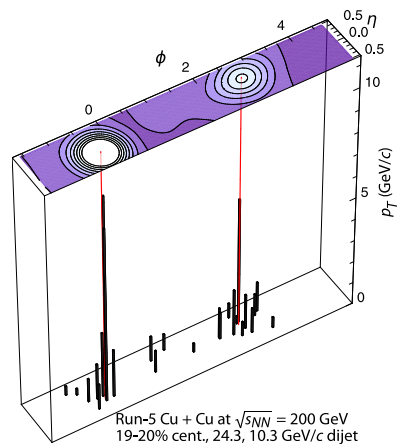}}
  \caption{A PHENIX Run-5 \CuCu{} at \sNNtwohundred{} dijet event at
    $\approx 20\%$ centrality. Charged tracks and photons are shown at
    the bottom by a Lego plot. The distribution of filter output
    values of the event is shown at the top as a contour plot. The
    maxima in the filter density are reconstructed as jet axes, shown
    as red lines at the positions on the contour and Lego plots.}
  \label{fig:run_5_cu_cu_event_display_back_0}
\end{figure}

Fluctuations in the underlying event of heavy ion collisions are known
to cause a false apparent jet production, if a jet reconstruction
algorithm is applied without proper handling of the
background~\cite{Stewart:1990wa}. At the collision energy of RHIC, the
intrinsic jet production rate is far below the high-\pT{} tail of the
background fluctuation across a large \pT{} range that is
statistically accessible to RHIC. Based on the \pp{} jet
cross-sections presented later in this paper, we estimate that for the
central \CuCu{} at \sNNtwohundred{} and $p_T$ between
$10$--$20\,\mathrm{GeV}/c$ and $20$--$30\,\mathrm{GeV}/c$, the jet
yields are $N_\mathrm{evt}^{-1} dN/dy \approx 4\times 10^{-3}$ and
$8\times 10^{-5}$, respectively, which means that rare fluctuations
can contribute significantly to the jet yield, if the background
cannot be suppressed below this level. Avoiding the $p_T$ range with
background contribution would require measuring jets only at the very
end of the $p_T$ range, where the statistics are poor. The presence of
rare fluctuations also suggest that a simple removal of low-\pT{}
particles would only sample the stronger fluctuations in the rare,
high-\pT{} tail, while biasing both the fragmentation and energy of
the reconstructed jets.

We therefore developed a fake jet rejection strategy that is based on
the jet versus background shape and can achieve a higher rejection
rate than previously proposed algorithms for the LHC
(e.g.~\cite{Grau:2008ed}). As will be demonstrated below, the fake
rejection provides a fast rise to unity efficiency within the RHIC
accessible \pT{} range.

\section{Jet reconstruction by Gaussian filtering}

It can be shown that the iterative cone algorithm is equivalent to
finding local maxima of a filter output in $(\eta, \phi)$ with a flat
angular weighting $k(r^2) = \theta(R^2 - r^2)$ with $r^2 = \eta^2 +
\phi^2$ (note that unlike $k(r^2)$, the filter kernel $h(r^2) \propto
-\int dr^2 k(r^2) \propto \max(0, 1 - r^2 / R^2)$, and not
flat)~\cite{Cheng:1995ms,Fashing:2005ms}. The cone algorithm entails a
specific choice of angular weighting. The Gaussian filter is based on
another, which takes advantage of jets being a collimated emission of
particles, and enhances the center of the jet and suppresses the
possible contribution from the event background in the periphery.
Expressed in the filter form described below, the algorithm samples
the entire possible $(\eta, \phi)$ range and is seedless. By
additionally avoiding a sharp radial cutoff, the algorithm therefore
becomes analytically collinear and infrared safe (we also verified the
practical infrared safety using a procedure similar
to~\cite{Salam:2007xv}).

A combined event transverse momentum density that contains both the
final state particles $p_{T,i}$, and $p_T^\mathrm{bg}(\eta, \phi)$,
which represents an independently evaluated average contribution from
the underlying event, can be defined as
\begin{equation}
  p_T(\eta, \phi) = \sum_{i \in F} p_{T,i} \delta(\eta - \eta_i)
  \delta(\phi - \phi_i) - p_T^\mathrm{bg}(\eta, \phi),
  \label{eq:perp_density}
\end{equation}
The Gaussian filtering of \pT{} is the linear-circular convolution of
$p_T(\eta, \phi)$ with a Gaussian distribution
\begin{equation}
  p_T^\mathrm{filt}\!(\eta, \phi) =
  \!\int\!\!\!\!\int\! d\eta' d\phi'
  p_T(\eta', \phi') e^{-((\eta - \eta')^2 + (\phi -
    \phi')^2)/2\sigma}.
\end{equation}
The output of the filter for a given $(\eta, \phi)$ position is the
Gaussian-weighted transverse momentum in that event above the average
background from the underlying event. The local maxima in
$p_T^\mathrm{filt}(\eta, \phi)$ are the reconstructed jets using the
Gaussian filter.

Figures \ref{fig:run_5_p_p_event_display_back_0}--%
\ref{fig:run_5_cu_cu_event_display_back_0} demonstrate the behavior
for a \pp{} and \CuCu{} event, respectively. Charged tracks and
photons are shown at the bottom by a Lego plot. The distribution of
filter output values of the event is shown at the top as a contour
plot. The maxima in the filter density are reconstructed as jet axes,
shown as red lines at the positions on the contour and Lego plots.

\begin{figure}[t]
  \centerline{\includegraphics[width=3.125in]{%
      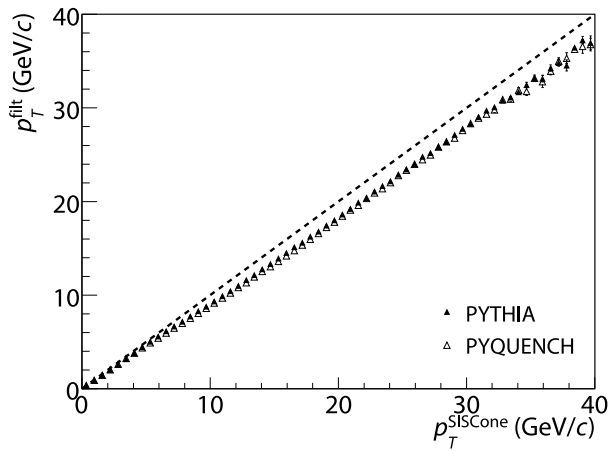}}
  \caption{Comparison of the jet energy scale between the $\sigma =
    0.3$ Gaussian filter ($p_T^\mathrm{filt}$) and the $R = 0.4$
    SISCone ($p_T^\mathrm{SISCone}$, with $0.5$ overlap threshold)
    \cite{Salam:2007xv} for \pp{} collisions at \stwohundred{} using
    \textsc{pythia} \cite{Sjostrand:2006za} and central \AuAu{}
    collisions at \sNNtwohundred{} using \textsc{pyquench}
    \cite{Lokhtin:2005px} at the event generator level (no detector
    effects). The dashed line indicates the position of
    $p_T^\mathrm{filt} = p_T^\mathrm{SISCone}$.}
  \label{fig:energy_scale_filter_siscone_pythia_pyquench_profile}
\end{figure}

The angular weighting of the Gaussian filter modifies the energy
summation and therefore will not produce the same jet energy as e.g.\
the cone algorithm. To evaluate the difference between the energy
scales of the $\sigma = 0.3$ Gaussian filter and the $R = 0.4$ SISCone
algorithms ($0.5$ overlap threshold)~\cite{Salam:2007xv}, we show in
Figure \ref{fig:energy_scale_filter_siscone_pythia_pyquench_profile}
the jet energy scale comparison for both \pp{} collisions at
\stwohundred{} using \textsc{pythia}~\cite{Sjostrand:2006za} and
central \AuAu{} collisions at \sNNtwohundred{} using
\textsc{pyquench}~\cite{Lokhtin:2005px} at the event generator level.
Further study of its property in \pp{} events are detailed in
\cite{Lai:2008zp}.

In heavy ion events, the background contribution $p_T^\mathrm{bg}$
will depend on the collision centrality and reaction plane. The
Gaussian filter allows these collision variables to be fully
parameterized, which is crucial for heavy ion jet reconstruction
algorithm. Due to the narrow pseudorapidity coverage at PHENIX, we
also parametrize $p_T^\mathrm{bg}$ depending on the collision vertex
$z$ position. Since in the \CuCu{} system at \sNNtwohundred{}, the
event-by-event fluctuation strongly dominates over flow effects, we do
not parametrize with respect to the reaction plane (note that the
linearity in the jet definition means that a weak flow simply
translates into an additional \pT{} smearing).

\section{Experimental setup}

\begin{figure}[t]
  \centerline{\includegraphics[width=2.75in]{%
      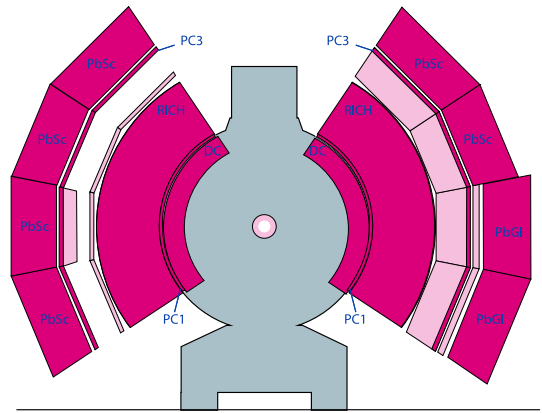}}
  \caption{The PHENIX central arm detectors for RHIC Run-5 (year
    2004/2005), viewed along the beam axis from the south towards
    north. Dark regions indicate detectors used for the jet
    reconstruction: The drift chamber (DC), the pad chamber layers 1
    and 3 (PC1/PC3), the ring-imaging \v{C}erenkov detector (RICH),
    and the Pb scintillator (PbSc) and Pb glass (PbGl) electromagnetic
    calorimeters.}
  \label{fig:central_arm_run_4_jet_reconstruction}
\end{figure}

Figure \ref{fig:central_arm_run_4_jet_reconstruction} shows the Run-5
PHENIX central arm configuration for RHIC Run-5 (year 2004/2005). The
central arm detectors used for jet reconstruction are the drift
chamber (DC), the pad chamber layers 1 and 3 (PC1/PC3), the
ring-imaging \v{C}erenkov detector (RICH), and the electromagnetic
calorimeters (EMCal). For the data presented in this paper, DC/PC1/PC3
provide momentum measurement for charged particles, and the EMCal the
energy for photons and electrons. Two calorimeter technologies were
used, 6 of the total 8 sectors are covered by Pb-scintillator (PbSc)
EMCal, 2 sectors by Pb-glass (PbGl) calorimeters.

Pattern recognition and momentum reconstruction of the tracking system
is performed using a combinatorial Hough transform. The $p_T$ scale is
determined by time-of-flight measurement of $\pi^\pm$, $K^\pm$, and
$p/\bar{p}$.

Since PHENIX currently does not have the ability of performing
in-field tracking\footnote{a Si vertex detector upgrade is going to
  provide such a capability and is scheduled to be installed in year
  2010, see e.g.\ \cite{Heuser:2003vd}}, conversion electrons in the
DC can produce a displaced track that appears to the momentum
reconstruction to be a very high-\pT{} track originating from the
event vertex. Information from the RICH and the $dE/dx$ measurement is
therefore used to identify and remove these conversion electrons. To
provide additional suppression for jets with $p_T^\mathrm{rec} >
20\,\mathrm{GeV}/c$, we use the fact that conversion electrons are
unlikely to angularly coincide with a high-\pT{} jet, and require the
reconstructed jet to have a minimum jet multiplicity of 3 particles,
and the charged fraction of the jet \pT{} to be below $0.9$ to remove
events with single track and falsely large $p_T$ values.

Shower shape cuts were applied to the EMCal clusters to
remove clusters generated by hadronic showers. The absolute energy
scale of the calorimeter clusters is determined both by the
reconstructed $\pi^0$ masses from $\pi^0 \rightarrow \gamma\gamma$
decay, and checked by comparing RICH identified $e^\pm$ momenta from
tracking against the corresponding cluster energies. The residual
uncertainty in the energy scale is $\pm 3\%\text{ (syst.)}$.

The PHENIX minimum bias (MB) trigger is defined by the coincident
firing of the two beam-beam counters (BBC) located at $3.0 < \eta <
3.9$. The Run-5 BBC cross section is $\sigma_\mathrm{BBC} = 22.9\pm
2.3\,\mathrm{mb}\text{ (syst.)}$, measured using the Van de
Meer/vernier scan method. The efficiency of BBC firing on an event
containing a jet with $p_T^\mathrm{rec} > 2\,\mathrm{GeV}/c$ is
$\epsilon_\mathrm{BBC} = 0.86\pm 0.05\text{ (syst.)}$ and within that
uncertainty, approximately constant with respect to
$p_T^\mathrm{rec}$. For both our \pp{} and \CuCu{} measurements, we
require the collision vertex to be within $|z| < 25\,\mathrm{cm}$
along the beam axis, derived from the BBC timing information.

All the charge tracks and electromagnetic clusters in the EMCal
passing the described cuts are used in the Gaussian filter, and all
resulting maxima are considered candidate jets. Since we do not
explicitly split jet pairs with small angular separation, it is
possible to reconstruct jet pairs with substantial overlap. In reality
however, we rarely observe jets reconstructed with an angular
separation of $\Delta R < 0.5$.

\section{Jet spectrum in \pp{}}

The data presented in this section were obtained from the PHENIX \pp{}
dataset from the RHIC Run-5 (year 2004/2005). After removal of bad
quality runs, a total of $1.47\times 10^9$ minimum bias \pp{} and
$1.16\times 10^9$ triggered \pp{} events are being used.

PHENIX can trigger on high-\pT{} and electromagnetic processes using
the central arm EMCal-RICH-trigger (ERT). For the result presented
here, we use a trigger that requires a total energy $E >
1.4\,\mathrm{GeV}$ deposited in a $4\times 4$ group of calorimeter
towers. This trigger is well-suited for jet measurement due to its low
noise level and fast efficiency saturation with respect to the jet
\pT{}. The efficiency is evaluated in term of the $p_T^\mathrm{rec}$
of the most energetic jet in the event, and rises from approximately
$0.25$ for $p_T^\mathrm{rec} \approx 2.2\,\mathrm{GeV}/c$ to $0.9$ for
$p_T^\mathrm{rec} \approx 9.5\,\mathrm{GeV}/c$. After correction for
the efficiency, the minimum bias and triggered datasets agree within
an uncertainty of $5\%\text{ (syst.)}$ in the range from
$p_T^\mathrm{rec} = 2\,\mathrm{GeV}/c$ up to $30\,\mathrm{GeV}/c$
(limited by the minimum bias data set statistics). Therefore, we
calculate a combined jet yield per minimum bias event according to
\begin{equation}
  \frac{1}{N_\mathrm{evt}} \frac{dN}{dp_T^{pp}} =
  \frac{1}{N_\mathrm{MB}}
  \frac{\epsilon_\mathrm{ERT}^{-1}(p_T^{pp})
    \frac{dN_\mathrm{ERT}}{dp_T^{pp}} +
    \frac{dN_\mathrm{evt}}{dp_T^{pp}}}{s_\mathrm{MB} + 1},
\end{equation}
where $s_{MB} = 37.34$ is the average scale down applied to the
minimum bias trigger.

The spectrum is then unfolded by the regularized inversion of the
Fredholm equation
\begin{equation}
  \frac{dN}{dp_T^{pp}} = \int dp_T P(p_T^{pp} | p_T) \frac{dN}{dp_T}
  \label{eq:fredholm}
\end{equation}
using singular value decomposition (SVD). The simultaneous
minimization of the second order finite-differences is use as a
constraint to the continuity of the unfolding result. This procedure
is implemented in the software package \textsc{guru}
\cite{Hoecker:1996sa}. The transfer matrix of $P(p_T^{pp}
| p_T)$ expresses the (conditional) probability that a jet with the
true transverse momentum, \pT{}, is reconstructed with
$p_T^{pp}$.

\begin{figure}[t]
  \centerline{\includegraphics[width=3.125in]{%
      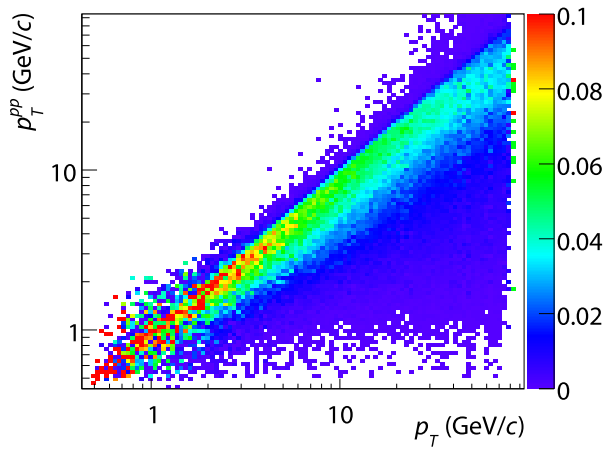}}
  \caption{The PHENIX jet $P(p_T^{pp} | p_T)$ transfer matrix for
    \stwohundred{} and $\sigma = 0.3$ Gaussian filter, derived from
    the \textsc{geant} simulation of $\approx 1.6\times 10^7$
    \textsc{pythia} events. The $p_T^{pp} < p_T$ region is dominated
    by $n$, $K_L^0$ energy loss.}
  \label{fig:p_p_energy_scale}
\end{figure}

We used \textsc{pythia} 6.4.20 with the (6.4.20-default) ``old''
multiparton interaction scheme and \textsc{geant} simulation to
evaluated $P(p_T^{pp} | p_T)$. A total of $1.6\times 10^7$ events were
simulated with 14 different minimum $Q^2$ settings varying
between$\sqrt{Q^2} > 0.5\,\mathrm{GeV}/c$ and $\sqrt{Q^2} >
64\,\mathrm{GeV}/c$. The transfer matrix $P(p_T^{pp} | p_T)$ resulting
from the simulation procedure is shown in figure
\ref{fig:p_p_energy_scale}. The $p_T^{pp} < p_T$ region is dominated
by $n$, $K_L^0$ energy loss.

We measured the \stwohundred{} \pp{} spectrum using the combined
PHENIX Run-5 minimum bias and triggered data. We require that the
reconstructed jet is at least by an angular distance of $d \ge
0.05\,\mathrm{rad}$ inside the infinite momentum PHENIX central arm
acceptance. The spectrum with respect to the true $p_T$ is then given
by
\begin{equation}
  \frac{E d^3\sigma}{dp^3} = \frac{1}{2\pi p_T} \frac{d^2\sigma}{dp_T
    dy} = \frac{\sigma_\mathrm{BBC}}{A\,\epsilon_\mathrm{BBC}}
  \frac{1}{p_T} \frac{1}{N_\mathrm{evt}} \frac{dN}{dp_T}
  \label{eq:sigma_pp}
\end{equation}
where $dN/dp_T$ is the unfolding result from \eqref{eq:fredholm} and
\begin{equation}
  A = 2 (\Delta\eta - 2 d) (\Delta\phi / 2 - 2 d)
  \label{eq:fiducial_area}
\end{equation}
the fiducially reduced PHENIX central arm acceptance area.

The regularized least square unfolding involves the regularization
parameter $\tau$ (or sometimes $\lambda^2$) and its choice translates
into an uncertainty on the global shape/low frequency component of a
jet spectrum. We evaluated the systematic uncertainty in the unfolded
spectrum due to such variation in the regularization parameter by
varying $\tau$ over the entire meaningful range between $\approx 4$
degrees of freedom up to the Nyquist frequency. We combine the
resulting, point-by-point estimate of the systematic uncertainty for
the spectrum shape as part of the total experimental systematic
uncertainty. The so evaluated systematic uncertainty is therefore
representative of the full range of regularization parameter choices.

The residual $\pm 3\%$ systematic uncertainty in the energy scale
translates into a constant uncertainty of $\pm 15\%\text{ (syst.)}$
for the jet spectrum (due to its power-law shape with largely constant
exponent).

\begin{figure}[t]
  \centerline{\includegraphics[width=3.125in]{%
      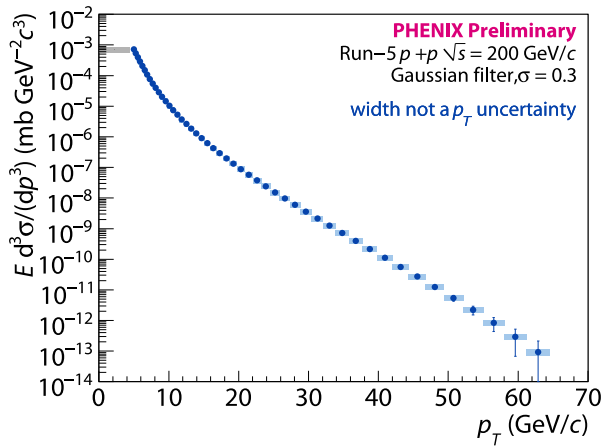}}
  \caption{PHENIX Run-5 \pp{} at \stwohundred{} invariant jet cross
    section spectrum as a function of $p_T$. The shaded box to the
    left indicates the overall normalization systematic uncertainty,
    shaded boxes associated with data points indicate point-to-point
    systematic uncertainties, and error bars indicate statistical
    uncertainties.}
  \label{fig:run_5_p_p_perp}
\end{figure}

Figure \ref{fig:run_5_p_p_perp} shows the PHENIX preliminary \pp{} jet
spectrum measured using the Gaussian filter by the procedures
described above, plotted in invariant cross sections. The shaded box
to the left indicates the overall normalization systematic
uncertainty, shaded boxes associated with data points indicate
point-to-point systematic uncertainties, and error bars indicate
statistical uncertainties. We show the unfolded spectrum out to the
$p_T$ bin where the nominal yield for the number of sampled events
reaches the level of $1$ jet, namely $60\,\mathrm{GeV}/c$.

\begin{figure}[t]
  \centerline{\includegraphics[width=3.125in]{%
      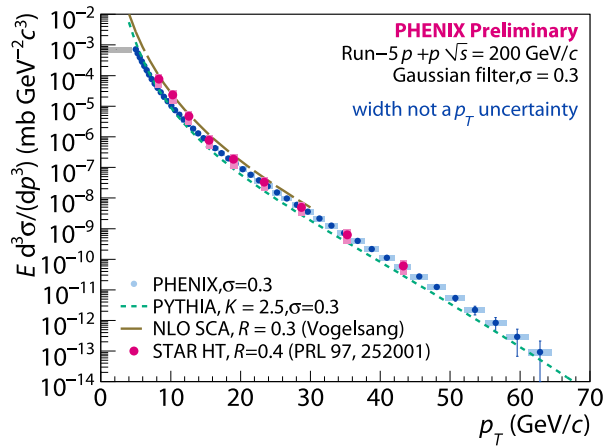}}
  \caption{PHENIX Run-5 \pp{} at \stwohundred{} invariant jet cross
    section spectrum as a function of $p_T$, with comparison to
    \cite{Abelev:2006uq}, next-to-leading order calculation from
    \cite{Jager:2004jh}, and \textsc{pythia} assuming $K = 2.5$. The
    shaded box to the left indicates the overall normalization
    systematic uncertainty, shaded boxes associated with data points
    indicate point-to-point systematic uncertainties, and error bars
    indicate statistical uncertainties.}
  \label{fig:run_5_p_p_perp_comparison_all}
\end{figure}

Figure \ref{fig:run_5_p_p_perp_comparison_all} shows the same spectrum
as in Figure \ref{fig:run_5_p_p_perp}, compared against the spectrum
from \cite{Abelev:2006uq}, the next-to-leading order (NLO) calculation
using the small cone approximation (SCA) \cite{Jager:2004jh}, and the
leading order \textsc{pythia} spectrum assuming $K = 2.5$. The
comparison to \cite{Abelev:2006uq} and NLO SCA involve different jet
definitions, a residual difference should be expected, even though for
$p_T > 15\,\mathrm{GeV}/c$ it appears to be small between filter and
cone jets for the Gaussian size $\sigma = 0.3$ used in this analysis.
Our spectrum is close to \cite{Abelev:2006uq} within its \pT{} reach.
The spectrum also follows approximately the shape of the NLO SCA
calculation, and the leading order \textsc{pythia} spectrum, if $K =
2.5$ is assumed. However, a more appropriate comparison would involve
Gaussian filter based NLO calculations, which we plan to perform in
the future.

\section{\CuCu{} results}

The data presented in this section were obtained from the PHENIX
\CuCu{} dataset from the RHIC Run-5 (year 2004/2005). The \pp{} data
presented above provide a baseline for the \CuCu{} measurements and
explicitly appear in the jet nuclear modification factor. After
removal of bad quality runs, a total of $1.58\times 10^8$ minimum bias
are used in the \CuCu{} analysis, covering the centrality range
$0$--$95\%$. Because of poor statistics and large uncertainties in the
$T_{AB}$ for very peripheral events, we excluded events in the
$80$--$95\%$ centrality range and divide the remainder into 4 bins.

The $p_T^\mathrm{bg}$ in \eqref{eq:perp_density} is parametrized as a
product of the centrality and vertex dependent total event
$p_T^\mathrm{tot}$ times a vertex dependent $p_T(\eta,
\phi)/p_T^\mathrm{tot}$ distribution. The values are estimated by
averaging over the minimum bias \CuCu{} events.

In heavy ion collisions, centrality-dependent fluctuations in the
underlying event will broaden the energy resolution of the jet
measurement beyond the detector response described in Section 4. The
differences between the \CuCu{} jet \pT{} scale, $p_T^\mathrm{CuCu}$
and the \pp{} energy scale must be accounted for in any comparison of
the jet spectra or evaluation of \RAA{}.

We evaluate the effects of the \CuCu{} underlying event on jet
measurements by embedding PHENIX-measured \pp{} events into the
minimum bias \CuCu{} events by combining their list of tracks and
clusters. We reconstruct the embedded events and evaluate the energy
and angular resolution and the reconstruction efficiency for jets that
match the input \pp{} jets within a radial separation $\Delta R <
0.3$. Both events are required to fall into the same $\Delta z =
5\,\mathrm{cm}$ bin in vertex position. The evaluated transfer matrix
is then recombined into the final centrality binning using
$N_\mathrm{coll}$ scaling. Due to the linearity of Gaussian filter,
for $p_T^{pp} > 8\,\mathrm{GeV}/c$ the energy smearing is constant
within our statistical uncertainty. We use this feature of the energy
response to extend the transfer matrix to higher \pT{} where the
limited statistics in the \pp{} sample would otherwise introduce
fluctuations into the transfer matrix that would have a negative
impact on the unfolding.

\begin{figure}[t]
  \centerline{\includegraphics[width=3.125in]{%
      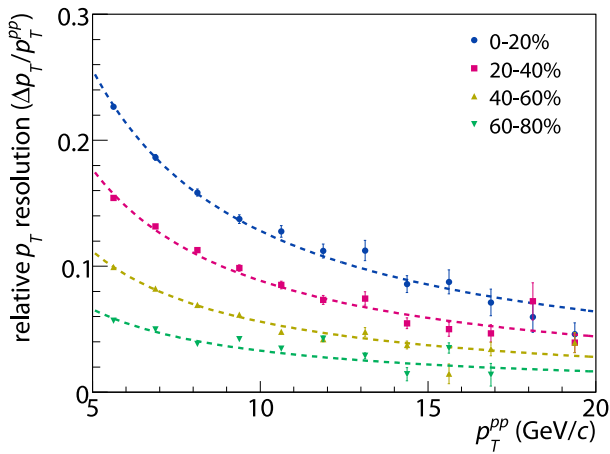}}
  \caption{The $\sigma = 0.3$ Gaussian filter \pT{} resolution with
    $\Delta p_T/p_T \propto 1/p_T$ (constant energy smearing) fits for
    different centralities of \CuCu{} collisions at \sNNtwohundred{},
    evaluated by embedding PHENIX Run-5 \pp{} events into PHENIX Run-5
    \CuCu{} events. The constant energy smearing behavior is expected
    for an (asymptotically) linear jet reconstruction algorithm.}
  \label{fig:run_5_cu_cu_perp_resolution_fit}
\end{figure}

Figure \ref{fig:run_5_cu_cu_perp_resolution_fit} shows the $\sigma =
0.3$ Gaussian filter \pT{} resolution for different centralities of
\CuCu{} collisions at \sNNtwohundred{}, evaluated by embedding PHENIX
Run-5 \pp{} events into PHENIX Run-5 \CuCu{} events. Superimposed are
$\Delta p_T/p_T \propto 1/p_T$ fits that shows the constant energy
smearing, which is expected for an (asymptotically) linear jet
reconstruction algorithm such as the Gaussian filter.

Due to the large $N_\mathrm{coll}$ in central \CuCu{} events, the
underlying event has potentially a high yield to contain an intrinsic
jet. This is avoided by requiring that reconstructed jet after
embedding matches a \pp{} jet with $p_T^{pp} > 4\,\mathrm{GeV}/c$
within an angular range of $\Delta R < 0.3$. The residual
contamination was evaluated by reconstructing the underlying event jet
spectrum within $\Delta R$ of the original \pp{} jet axis, and is
found to be $< 10^{-2}$.

In the measurement presented here, we use two approaches to correct
for the energy scale difference in \CuCu{} collisions: (a) by
unfolding the \CuCu{} spectrum using the transfer matrix
$P(p_T^\mathrm{CuCu}|p_T^{pp})$, the \RAA{} is then calculated by
comparing against the \pp{} spectrum; and (b) by embedding \pp{}
events into \CuCu{} events, so that the \pp{} spectra attains the
\CuCu{} background induced energy smearing, and the \RAA{} is derived
by comparing the \CuCu{} spectrum to the embedded spectrum.

\subsection{Fake jet rejection}

\begin{figure}[t]
  \centerline{%
    \includegraphics[width=1.5in]{%
      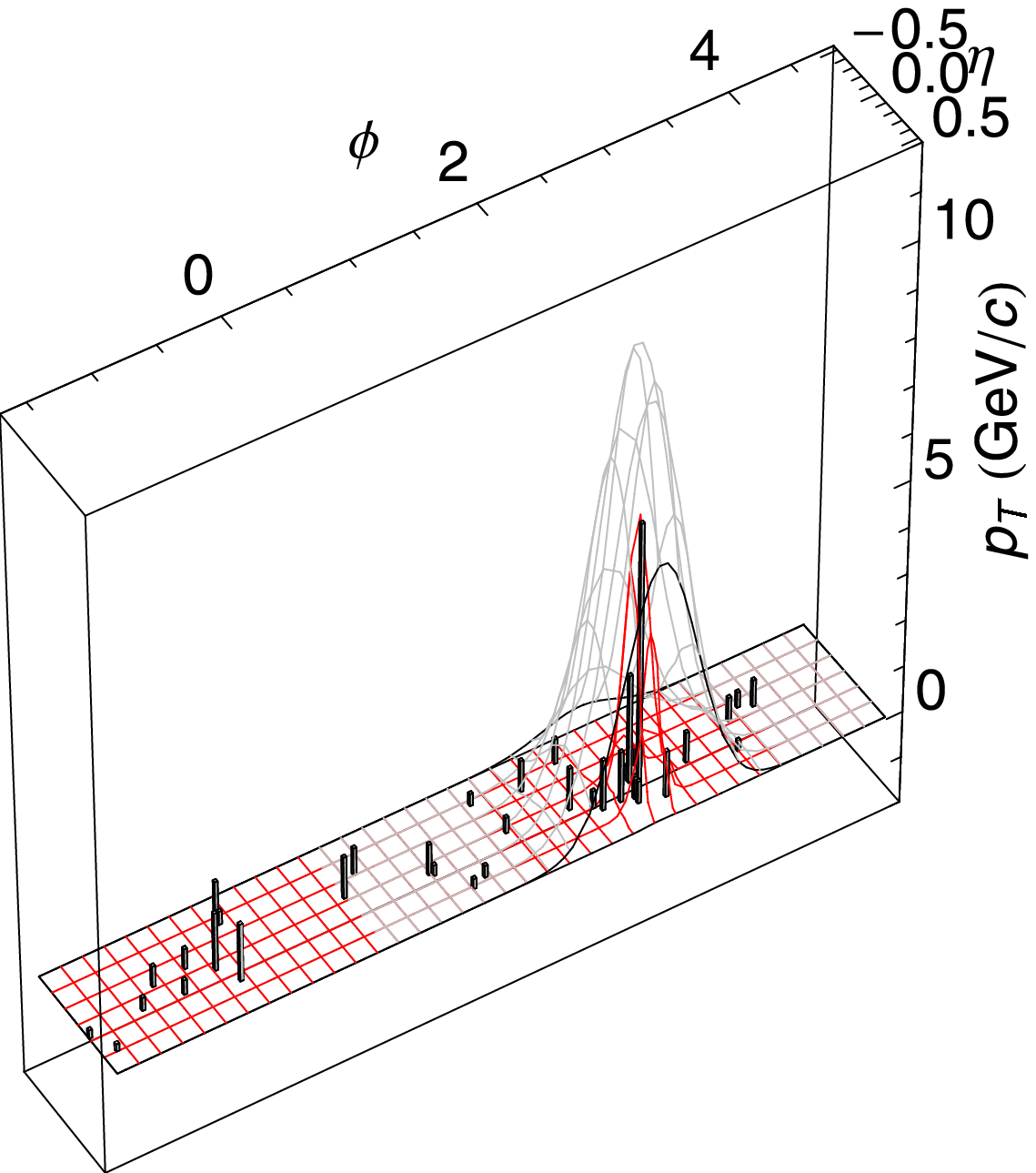}\hfill%
    \includegraphics[width=1.5in]{%
      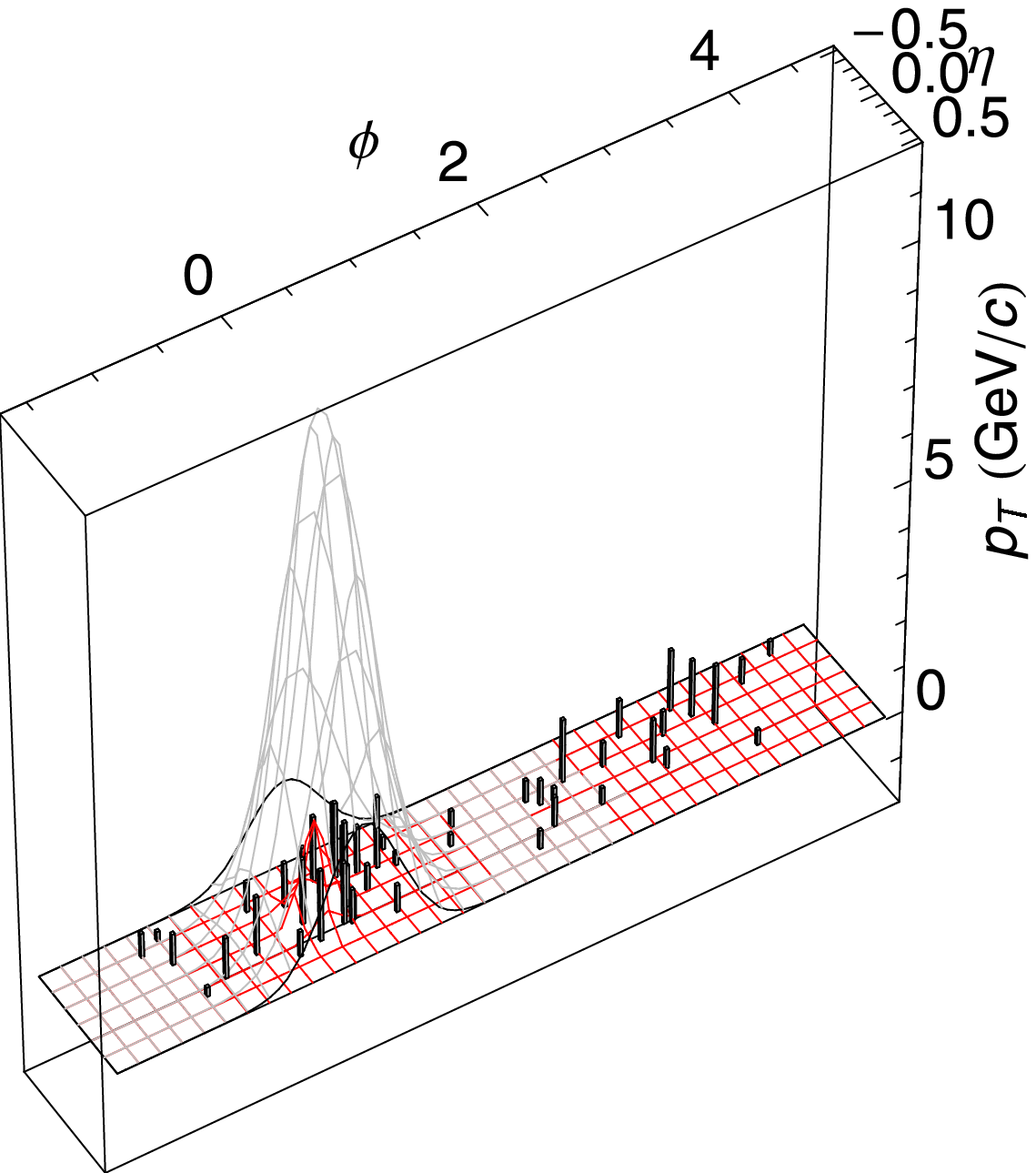}}
  \caption{Comparison between a PHENIX Run-5 \CuCu{} event with a
    $9.6\,\mathrm{GeV}/c$ jet passing fake rejection and an event with
    $10.8\,\mathrm{GeV}/c$ ``jet'' that fails the fake rejection, and
    is presumed to be a background fluctuation}
  \label{fig:event_display_fake_rejection}
\end{figure}

In order to reduce the contribution of fake jets, we apply a shape
based discriminant to the jets. The discriminant is defined as
\begin{equation}
  g_{\sigma_\mathrm{dis}}(\eta, \phi) = \sum_{i \in \mathrm{fragment}}
  p_{T,i}^2 e^{-((\eta_i - \eta)^2 + (\phi_i -
    \phi)^2)/2\sigma_\mathrm{dis}},
  \label{eq:discriminant}
\end{equation}
where $(\eta, \phi)$ is the reconstructed jet axis. The discriminant
size $\sigma_\mathrm{dis} = 0.1$ to chosen to be approximately or
below the size $\Delta R^\mathrm{bg} = \sqrt{2\pi/(dN/d\eta)}$, which
is the characteristic background particle separation. If the local
distribution of high-\pT{} particle indicates that the jet may be
misreconstructed in its angle (e.g.\ a background fluctuation beneath
the jet with a nonzero gradient), we also search for an $(\eta, \phi)$
where $g_{\sigma_\mathrm{dis}}$ might be maximized. We denote this
``adapted'' discriminant $g_{\sigma_\mathrm{dis}}'$. In
\textsc{hijing} studies, we found this discriminant, when applied to
central \AuAu{} collisions at RHIC, to significantly outperform
discriminants proposed for the LHC, such as the $p_T/\langle
A\rangle$~\cite{Cacciari:2007fd} or $\Sigma j_T$ \cite{Grau:2008ed}.

Figure \ref{fig:event_display_fake_rejection} illustrates the behavior
of the fake jet rejection using two actual PHENIX Run-5 \CuCu{}
events.

With the definition in \eqref{eq:discriminant}, it is also possible to
achieve a largely centrality independent efficiency turn-on. This
feature is important in order to constrain the systematic impact in a
centrality dependent measurement. We could, in principle, adapt the
fake rejection to the collision centrality but that by keeping the
rejection threshold fixed we obtain a nearly centrality independent
jet finding efficiency. Using the studies of the discriminant
distribution in \CuCu{} collisions and on the dijet $\Delta\phi$
analysis presented below, we have chosen a nominal discriminant
threshold of $g_{0.1}' > 17.8\,(\mathrm{GeV}/c)^2$.

There are several methods to investigate the sensitivity of
reconstructed jets to the fake rejection. The discriminant
distribution shows that for $p_T > 16\,\mathrm{GeV}/c$, the jets have
discriminant values that lie mostly above our \CuCu{} fake jet
rejection threshold of $g_{\sigma_\mathrm{dis}}' >
17.8\,(\mathrm{GeV}/c)^2$. Changing or removing the threshold
therefore has little impact on our spectra and \RAA{} above
$12$--$16\,\mathrm{GeV}/c$.

\begin{figure}[t]
  \centerline{\includegraphics[width=3.125in]{%
      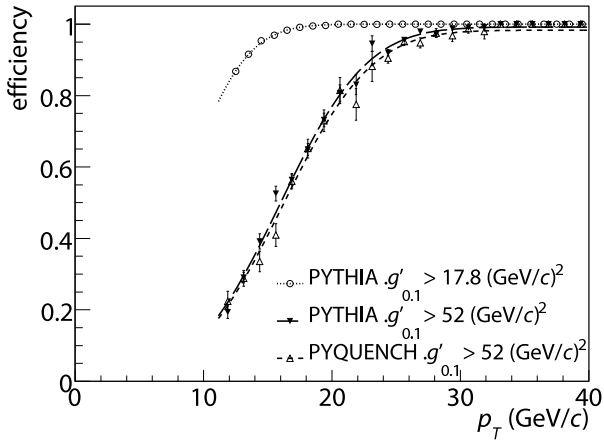}}
  \caption{Efficiency for the $g_{0.1}' > 52\,(\mathrm{GeV}/c)^2$ fake
    jet rejection that is suitable for central \AuAu{} collisions at
    \sNNtwohundred{} for \pp{} collisions at \stwohundred{} using
    \textsc{pythia} \cite{Sjostrand:2006za} and central \AuAu{}
    collisions at \sNNtwohundred{} using \textsc{pyquench}
    \cite{Lokhtin:2005px} at the event generator level (no detector
    effects), compared to the \textsc{pythia} efficiency for $g_{0.1}'
    > 17.8\,(\mathrm{GeV}/c)^2$ that is used for \CuCu{} collisions.}
  \label{fig:efficiency_pythia_pyquench}
\end{figure}

\begin{figure}[t]
  \centerline{\includegraphics[width=3.125in]{%
      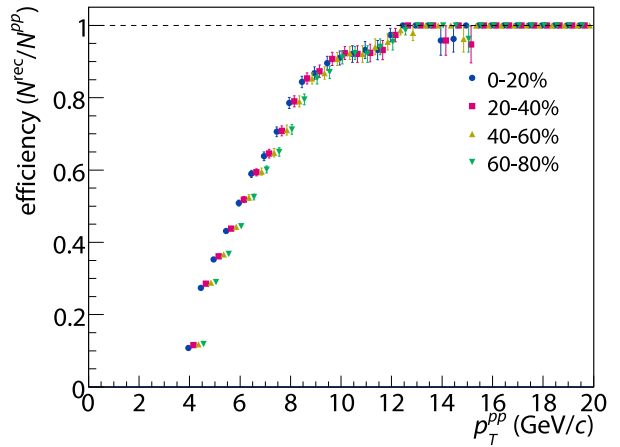}}
  \caption{Efficiency for the $g_{0.1}' > 17.8\,(\mathrm{GeV}/c)^2$
    fake jet rejection for different centralities of \CuCu{}
    collisions at \sNNtwohundred{}, evaluated by embedding PHENIX
    Run-5 \pp{} events into PHENIX Run-5 \CuCu{} events}
  \label{fig:run_5_cu_cu_efficiency}
\end{figure}

We further studied the sensitivity of both efficiency after fake jet
rejection and the energy scale using \textsc{pythia} and
\textsc{pyquench} at the fake jet rejection level $g_{0.1}' >
52\,(\mathrm{GeV}/c)^2$ required for central \AuAu{} collisions at
\sNNtwohundred{}. Both the efficiency and the jet energy scale
compared to SISCone shows very small modification, i.e.\ below the
present systematic uncertainty in the overall energy scale. This is
shown in Figure \ref{fig:efficiency_pythia_pyquench}, and the
comparison between \textsc{pythia} and \textsc{pyquench} jet energy
scales was shown in Figure
\ref{fig:energy_scale_filter_siscone_pythia_pyquench_profile}. Figure
\ref{fig:run_5_cu_cu_efficiency} shows the efficiency for the
$g_{0.1}' > 17.8\,(\mathrm{GeV}/c)^2$ fake rejection for different
centralities of \CuCu{} collisions at \sNNtwohundred{}.

\begin{figure}[t]
  \centerline{\includegraphics[width=3.125in]{%
      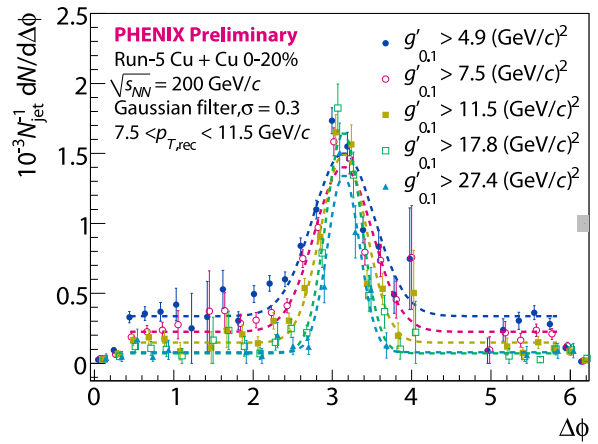}}
  \caption{Run-5 \CuCu{} at \sNNtwohundred{} azimuthal correlation for
    $0$--$20\%$ centrality dijets in yields and with different
    $g_{0.1}'$ fake jet rejection thresholds. The nominal fake jet
    rejection threshold used is $g_{0.1}' >
    17.8\,(\mathrm{GeV}/c)^2$.}
  \label{fig:run_5_cu_cu_dphi_fr}
\end{figure}

Angular correlation of dijets can be used to study the yield of
residual fake jets. The yield of three and more jets at large angle
with respect the leading dijet axis (e.g.\ due to multiple collisions
or initial and final state radiation) is strongly suppressed at
high-$p_T$, and therefore the decorrelated yield approximately
perpendicular to the dijet axis is a good estimator of the residual
contamination by fake jets. In the \CuCu{} data, the $\Delta\phi$
distribution saturates at our normal discriminant threshold of
$g_{0.1}' > 17.8\,(\mathrm{GeV}/c)^2$. At $p_T^\mathrm{CuCu} \approx
7.5\,\mathrm{GeV}/c$ for $0$--$20\%$ centrality, this translates into
an upper bound of $10\%$ for the fake jet contamination in the jet
yield. Figure \ref{fig:run_5_cu_cu_dphi_fr} shows the effect of the
increasing fake jet rejection on the $\Delta\phi$ distribution using
the $0$--$20\%$ centrality Run-5 \CuCu{} data.

\subsection{\CuCu{} jet spectra and nuclear modification factors}

We unfold the raw spectrum $dN/dp_T^\mathrm{CuCu}$ using the
centrality dependent transfer matrix $P(p_T^\mathrm{CuCu}|p_T^{pp})$
and obtain the spectrum in term of the $dp_T^{pp}$ energy scale,
$dN/dp_T^{pp}$. The invariant yield in the respective $p_T$ scale is
then given by
\begin{equation}
  \frac{1}{N_\mathrm{evt}} \frac{E d^3N}{dp^3} =
  \frac{1}{\epsilon_\mathrm{fr}(p_T)} \frac{1}{A} \frac{1}{p_T}
  \frac{1}{N_\mathrm{evt}} \frac{dN}{dp_T},
\end{equation}
with the same fiducially reduced area $A$ defined in
\eqref{eq:fiducial_area}, and where $\epsilon_\mathrm{fr}$ is the fake
rejection efficiency (compare figure
\ref{fig:run_5_cu_cu_efficiency}). For both the jet spectra and \RAA{}
measurement, we conservatively restrict our $p_T$ range to the region
with $\epsilon_\mathrm{fr} > 0.75$. When applying the charged fraction
cut in \CuCu{} events, we only include particles with $p_T >
1.5\,\mathrm{GeV}/c$, so that background fluctuations cannot cause us
to accept fake jets generated by late conversion electrons.

\begin{figure}[t]
  \includegraphics[width=3.125in]{%
    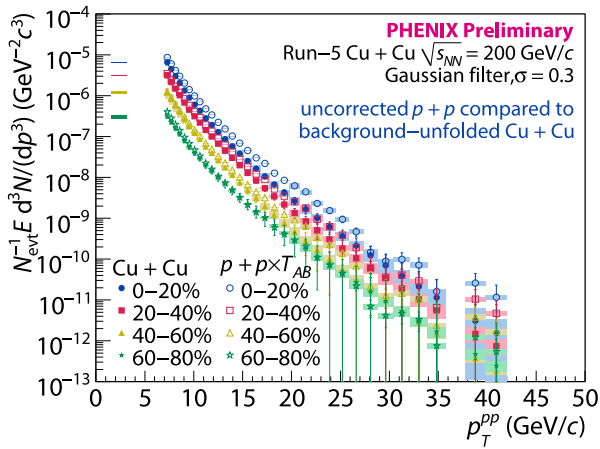}
  \caption{PHENIX Run-5 \CuCu{} at \sNNtwohundred{} invariant jet
    yield as function of $p_T^{pp}$, with comparison to the $\langle
    T_{AB}\rangle$ scaled \pp{} cross section. The shaded box to the
    left indicates the centrality-dependent systematic uncertainty in
    the normalization, shaded boxes associated with data points
    indicate point-to-point systematic uncertainties, and error bars
    indicate statistical uncertainties.}
  \label{fig:run_5_cu_cu_perp_unfolding}
\end{figure}

Figure \ref{fig:run_5_cu_cu_perp_unfolding} shows the invariant jet
yield as function of $p_T^{pp}$ together with the $\langle
T_{AB}\rangle$ scaled \pp{} cross section, where $T_{AB}$ is the
nuclear overlap function with \CuCu{} values given in
\cite{Adare:2008cx} are used. The shaded box to the left indicates the
centrality-dependent systematic uncertainty in the normalization,
shaded boxes associated with data points indicate point-to-point
systematic uncertainties, and error bars indicate statistical
uncertainties.

The nuclear modification factor is defined by
\begin{equation}
  R_{AA} = \frac{N_\mathrm{evt}^{-1}dN_\mathrm{CuCu}/dp_T}{\langle
    T_{AB}\rangle d\sigma_{pp}/dp_T}.
\end{equation}
For the \RAA{} in $p_T^{pp}$ energy scale and using unfolding, we
divide the spectra shown in Figure
\ref{fig:run_5_cu_cu_perp_unfolding} by the raw \pp{} spectrum in
$p_T^{pp}$.

To check for potential systematic errors resulting from the unfolding,
we alternatively evaluated the \RAA{} by comparing the raw \CuCu{}
spectrum to the embedded \pp{} spectrum. Since the extracted \RAA{} do
not show a significant \pT{} dependence within the uncertainties, we
can compare between the two sets of \RAA{} despite the difference in
the jet energy scale. We additionally studied the sensitivity of the
\RAA{} to vertex and fiducial cuts. The so evaluated, centrality
dependent differences from the unfolding/embedding comparison, vertex
dependence, fiducial dependence, are combined to obtain the centrality
dependent systematic uncertainties.

\begin{figure}[t]
  \includegraphics[width=3.125in]{%
    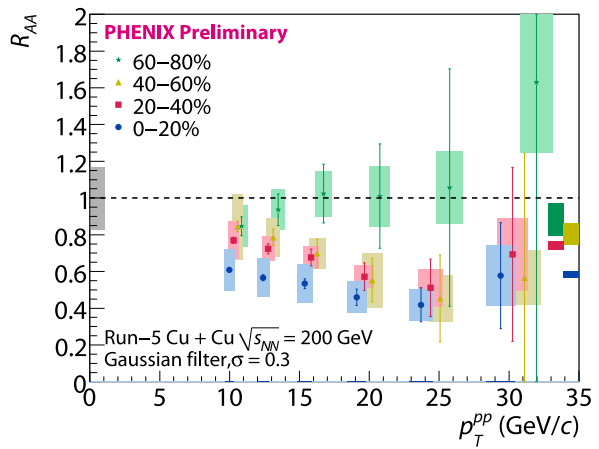}
  \caption{PHENIX Run-5 \CuCu{} at \sNNtwohundred{} \RAA{} derived
    from unfolding. The shaded box to the left indicates the
    \pp{}--\CuCu{} systematic uncertainty in the jet energy scale,
    shaded boxes to the right shows centrality dependent systematic
    uncertainty between embedding and unfolding, shaded boxes
    associated with data points indicate point-to-point systematic
    uncertainties, and error bars indicate statistical uncertainties.}
  \label{fig:run_5_cu_cu_r_aa_unfolding}
\end{figure}

\begin{figure*}[t]
  \centerline{%
    \includegraphics[width=3.125in]{%
      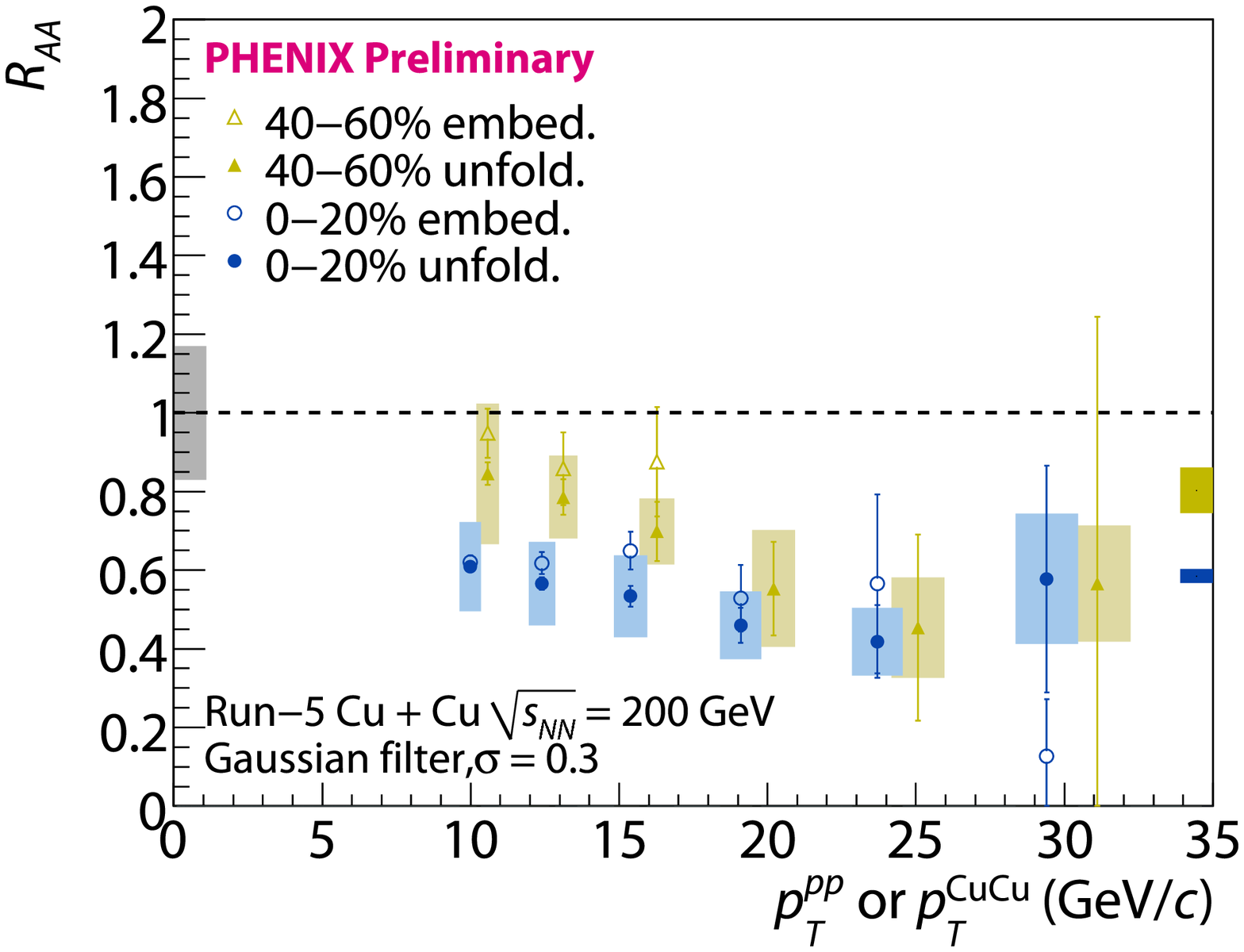}\hfill%
    \includegraphics[width=3.125in]{%
      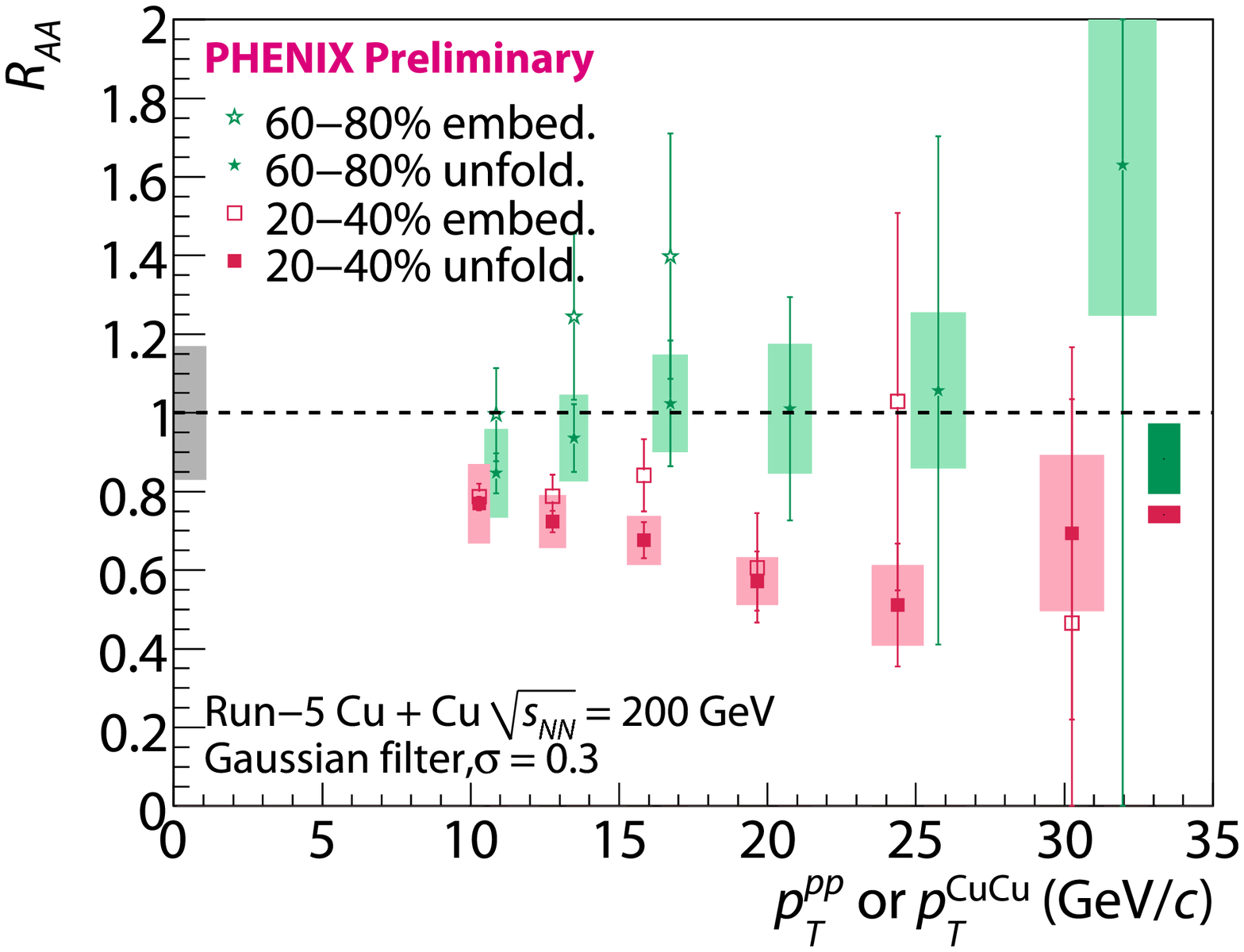}}
  \caption{Comparison between the PHENIX Run-5 \CuCu{} at
    \sNNtwohundred{} \RAA{} derived from unfolding (filled symbols)
    and embedding (open symbols). The shaded box to the left indicates
    the \pp{}--\CuCu{} systematic uncertainty in the jet energy scale,
    shaded boxes to the right shows centrality dependent systematic
    uncertainty between embedding and unfolding, shaded boxes
    associated with data points indicate point-to-point systematic
    uncertainties, and error bars indicate statistical uncertainties.
    Note that the flatness of \RAA{} makes a comparison across
    different energy scales possible.}
  \label{fig:run_5_cu_cu_r_aa_compare}
\end{figure*}

Figure \ref{fig:run_5_cu_cu_r_aa_unfolding} shows the extracted \RAA{}
using unfolding. The shaded box to the left indicates the
\pp{}--\CuCu{} systematic uncertainty in the jet energy scale, shaded
boxes to the right shows centrality dependent systematic uncertainty
between embedding and unfolding, shaded boxes associated with data
points indicate point-to-point systematic uncertainties, and error
bars indicate statistical uncertainties.

\begin{figure}[t]
  \includegraphics[width=3.125in]{%
    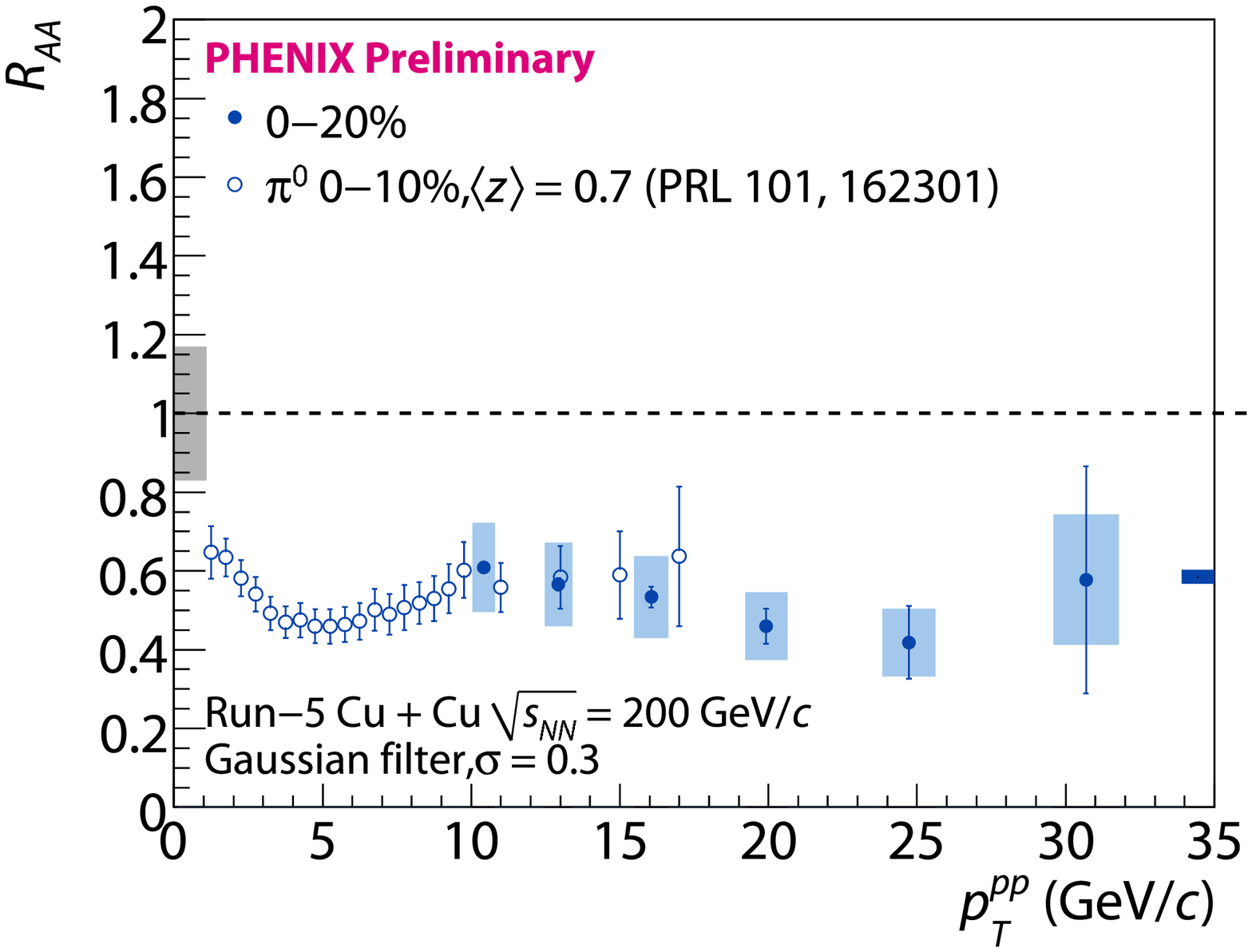}
  \caption{Comparison between the central PHENIX Run-5 \CuCu{} at
    \sNNtwohundred{} jet \RAA{} derived from unfolding and the $\pi^0$
    \RAA{}. The shaded box to the left indicates the \pp{}--\CuCu{}
    systematic uncertainty in the jet energy scale, shaded boxes to
    the right shows centrality dependent systematic uncertainty
    between embedding and unfolding, shaded boxes associated with data
    points indicate point-to-point systematic uncertainties, and error
    bars indicate statistical uncertainties. Note that while the
    flatness of \RAA{} makes a comparison across different energy
    scales possible, $\pi^0$ with $\langle z\rangle = 0.7$ has a
    different energy scale.}
  \label{fig:run_5_cu_cu_r_aa_unfolding_central}
\end{figure}

Figure \ref{fig:run_5_cu_cu_r_aa_compare} shows the comparison between
the \RAA{} derived using unfolding and embedding (with the same
notation as in Figure \ref{fig:run_5_cu_cu_r_aa_unfolding}). The level
of the suppression obtained with both methods is the same within the
statistical and unfolding-systematic uncertainties, which gives us
confidence that the unfolding procedure is not significantly biasing
the result. Figure \ref{fig:run_5_cu_cu_r_aa_unfolding_central}
compares the central $20\%$ suppression with the $\pi^0$ suppression
from \cite{Adare:2008cx} (with the same notation as in Figure
\ref{fig:run_5_cu_cu_r_aa_compare}). While the \RAA{} of $\pi^0$ has a
different energy scale than jets, both \RAA{} are approximately flat
with respect to \pT{} within our accessible range and therefore allows
a comparison.

We observe a \RAA{} that becomes gradually suppressed with increasing
centrality. The level of suppression in the most central $20\%$
centralities is at $R_{AA} \approx 0.5$--$0.6$ and comparable to that
of $\pi^0$.

\subsection{\CuCu{} jet-jet azimuthal correlations}

The \CuCu{} jet-jet azimuthal correlation is extracted by correcting
for the acceptance effect using the area-normalized mixed event yield
(e.g.\ \cite{Adler:2005ee}):
\begin{equation}
  \frac{dN(\Delta\phi)}{d\Delta\phi} = \frac{1}{A(\Delta\phi)}
  \frac{dN^\mathrm{raw}(\Delta\phi)}{d\Delta\phi}
\end{equation}
where $A(\Delta\phi)$ is the detector acceptance correction. Using a
Gaussian fit to the distribution, we extracted the width for
$7.5\,\mathrm{GeV}/c < p_T^\mathrm{CuCu} < 11.5\,\mathrm{GeV}/c$.
The widths are consistent within the uncertainty across all centrality
ranges.

\begin{figure}[t]
  \includegraphics[width=3.125in]{%
    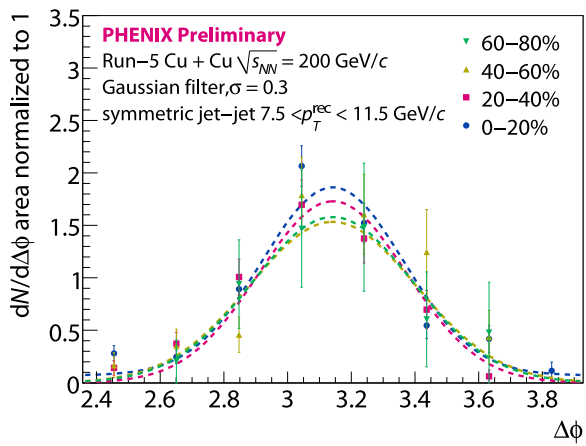}
  \caption{PHENIX Run-5 \CuCu{} at \sNNtwohundred{} azimuthal jet-jet
    correlation with Gaussian fits for jets with $7.5\,\mathrm{GeV}/c
    < p_T^\mathrm{CuCu} < 11.5\,\mathrm{GeV}/c$}
  \label{fig:run_5_cu_cu_dphi_zoomed}
\end{figure}

\begin{table}[t]
  \begin{center}
    \begin{tabular}{|l|c|c|c|}
      \hline \textbf{Centrality} & \textbf{Width}
      \\
      \hline 0--20\%  & $0.223\pm 0.017$\\
      \hline 20--40\% & $0.231\pm 0.016$\\
      \hline 40--60\% & $0.260\pm 0.059$\\
      \hline 60--80\% & $0.253\pm 0.055$\\
      \hline
    \end{tabular}
  \end{center}
  \caption{Widths of Gaussian fit to the PHENIX Run-5 \CuCu{} at
    \sNNtwohundred{} azimuthal
    angular correlation for jets with $7.5\,\mathrm{GeV}/c <
    p_T^\mathrm{CuCu} < 11.5\,\mathrm{GeV}/c$}
  \label{tab:run_5_cu_cu_dphi_fit}
\end{table}

Figure \ref{fig:run_5_cu_cu_dphi_zoomed} shows the azimuthal jet-jet
correlation with Gaussian fits for jets with $7.5\,\mathrm{GeV}/c <
p_T^\mathrm{CuCu} < 11.5\,\mathrm{GeV}/c$. Table
\ref{tab:run_5_cu_cu_dphi_fit} lists the Gaussian widths extracted
from the azimuthal jet-jet correlation for jets with
$7.5\,\mathrm{GeV}/c < p_T^\mathrm{CuCu} < 11.5\,\mathrm{GeV}/c$.

\section{Discussion}

Using the PHENIX Run-5 \pp{} dataset, we extracted the first RHIC
\pp{} spectrum that extends to $p_T \approx 60\,\mathrm{GeV}/c$ or $E
d^3\sigma/dp^3 \approx 0.3\,\mathrm{fb}\,\mathrm{GeV}^{-2} c^3$. This
effectively demonstrates the capability of PHENIX as a detector for
the study of jet physics.

Applying our algorithm to \CuCu{} collisions, were also able to
clearly demonstrate, for the first time, the feasibility of direct jet
reconstruction in heavy ion collisions and for all centralities. We
are able to show the centrality dependent onset of jet suppression
from peripheral to central collision. This demonstrates the capability
of the Gaussian filter algorithm as a heavy ion jet reconstruction
algorithm, that can be applied, fully unconstrained, in any heavy ion
collision setting at RHIC.

Our $\sigma = 0.3$ \RAA{} in \CuCu{} suggest that the jet production
is strongly suppressed, and is comparable to the suppression level of
high-\pT{} $\pi^0$. Strong suppression in reconstructed jets indicate
that significant amount of energy disappears from the angular region
covered by the size of the jet reconstruction algorithm. Since the
fake jet rejection has little impact on jets with $p_T >
12$--$16\,\mathrm{GeV}/c$, we observe an overall suppression that is
independent from the fake rejection scheme. However, if jets
significantly broadens at the presence of the medium, it is possible
that small angle jet reconstruction algorithm such as the $\sigma =
0.3$ Gaussian filter is deselecting these jets. We are current
pursuing jet spectra and \RAA{} evaluation with larger angle to
address this.

The azimuthal correlation suggest that the surviving parton traversing
the medium has very small transverse $k_T$ broadening.

While our current results are encouraging and demonstrates the
accessibility of a large inventory of heavy ion jet variables to
PHENIX, we would have to continue our measurement in term of
fragmentation properties and jet angular size dependence to understand
and complete our picture of the parton energy loss. Since our \pp{}
results probes in a previously unreached RHIC kinematic range,
additional comparisons with PQCD calculations could also provide
valuable insights.

\bibliography{dpf2009-ylai.bib}

\end{document}